\documentclass[reqno]{amsart}
\usepackage{hyperref}
\usepackage{amsmath,amssymb} 

\usepackage{euscript}
\usepackage[T2A]{fontenc}
\usepackage[utf8]{inputenc}
\usepackage{textcomp}

\allowdisplaybreaks

\theoremstyle{plain}
\newtheorem{theorem}{Theorem}[section]

\theoremstyle{definition}

\theoremstyle{remark}
\newtheorem{remark}[theorem]{Remark}
\numberwithin{equation}{section}

\begin{document}
	
	\title[ Disorder solutions for generalized 2D Ising Model with multi-spin interaction ]
	{Disorder solutions for generalized 2D Ising Model with multi-spin interaction		}
	
	\author[P. Khrapov]{Pavel Khrapov}
	\address{Pavel Khrapov, Department of Mathematics, Bauman Moscow State Technical University (5/1 2-nd
Baumanskaya St., Moscow 105005, Russia)
}  
	\email{khrapov@bmstu.ru }	
	


	\begin{abstract}
	For generalized 2D Ising model in an external magnetic field with the interaction of nearest neighbors, next nearest neighbors, all kinds of triple interactions and the quadruple interaction the formulas for finding free energy per lattice site in the thermodynamic limit were derived on a certain set of exact disordered solutions depending on seven parameters.
	 Lattice models are considered with boundary conditions with a shift (similar to helical ones), and with cyclic closure of the set of all points in natural ordering. The elementary transfer matrix with nonnegative matrix elements are constructed. On the set of disorder solutions the largest eigenvalue of the transfer matrix is constant for every size of considering planar lattice, and, in particular, in the thermodynamic limit. Free energy per lattice site in the thermodynamic limit is expressed through the natural logarithm of the largest eigenvalue of transfer matrix. This largest eigenvalue can be found for a special form of eigenvector with positive components. The numerical example show the existence of nontrivial solutions of the resulting systems of equations. The system of equations and the value of free energy in the thermodynamic limit will remain the same for 2D generalized Ising models with Hamiltonians, in which the values of two (out of four) neighboring maximal spins in the natural ordering are replaced by the values of the spins at any other two lattice points adjacent in the natural ordering, this significantly expands the set of models having disordered exact solutions. The high degree of symmetry and inductive construction of the components of the eigenvectors, which disappear when going beyond the framework of the obtained set of exact solutions, is an occasion to search for phase transitions in the vicinity of this set of disordered solutions.		
	\end{abstract}
	
	\maketitle

	\section{Introduction}\label{in}
	Ising model with interactions between pairs of nearest neighbours is one of the most studied systems in statistical mechanics. Some exact solutions ( with an analitical formulae for the partition function or free energy of the system) were obtained mainly for planar models, among these the exact Onsager's solution \cite{Onsager} of two-dimensional Ising model without an external magnetic field stands out clearly. 
	Publications \cite{Wu1982}-\cite{Aguilar_Braun}  are devoted to the theme of disorder solutions \cite{Stephenson},  obtained on a subset in the space of parameters of a physical system. Good reviews on this theme can be found in Wu F.Y.  \cite{Wu1982} , Baxter R.J.  \cite{baxter2016}.
	The transfer matrix apparatus is widely used in statistical physics \cite{Onsager}, \cite{Forrester_Baxter},  \cite{Minlos_Khrapov} - \cite{Khrapov3}, \cite{Khrapov4}- \cite{Khrapov5}. 	This article is a logical continuation of the author’s works  \cite{Khrapov3}, \cite{Khrapov4} and \cite{Khrapov5}, which outlines a general methodology for finding such disordered solutions for generalized Ising and Potts models, and explicitly some of these solutions for planar and three-dimensional generalized Ising model are obtained. 	 It is also shown  \cite{Khrapov3} how using the Levenberg -Marquardt method \cite{Levenberg} it is possible to obtain a numerical solution of the resulting systems of nonlinear equations.
	
	In this article, we consider planar generalised Ising models in the external magnetic field with general form of multi-spin interaction with nearest neighbors, next nearest neighbors, all kinds of triple interactions and the quadruple interaction. Toroidal boundary conditions are with a shift by one (similar to helical ones), and a cyclic closure of the set of all points (in natural ordering) \cite{Khrapov3}. For these models, elementary transfer matrices  \cite{Khrapov3} with non-negative elements were constructed, systems of equations were written and solved to find their largest eigenvalues (and the form of eigenvectors with positive components corresponding to these largest eigenvalues by the Perron-Frobenius theorem \cite{Perron}).
	 Thus, on the set of disorder solutions the largest eigenvalue of the transfer matrix is constant for every size of considering planar lattice, and, in particular, in the thermodynamic limit. The natural logarithm of the largest eigenvalue, taken with the opposite sign, is the division of the free energy and the product of Boltzmann's constant and temperature. The exact solutions of the system of equations, depending on seven parameters, were obtained in an explicit form. The system of equations and the value of free energy per lattice site in the thermodynamic limit remain the same for plane models with Hamiltonians in which the values of two (out of four) neighboring maximal spins in the natural ordering are replaced by the values of the spins at any other two lattice points neighboring in the natural ordering (\ref{v:rem1}), this is significantly expands the set of models having disordered exact solutions.
	 The periodicity of the eigenvectors’ components, which disappears when we go beyond the framework of 	the obtained set of disordered solutions, is a cause for searching 	phase transitions in the neighborhood of this set.
	
	Let us consider two-dimensional lattice (more detailed description of this lattice can be found in \cite{Khrapov3}, notations in this study are similar with notations in  \cite{Khrapov3}). Let us assume

	\begin{equation}\label{v:a}
\mathcal {L}_2=\{t=(t_1,t_2),t_i=0,1,...,L_i, i=1,2\}.
	\end{equation}
	Then let us make the following identification
	
	\begin{equation}\label{v:b}
	(L_1,t_2)\equiv (0,t_2+1), (L_1,L_2-1)\equiv (0,L_2) \equiv (0,0).
	\end{equation}
	
	Due to the process of  identifying points, the total number of lattice sites $\mathcal {L}_2 $    is equal to 
	$L=L_1  L_2$   . Thus, special boundary toroidal cyclic helical (with a shift) conditions are set on $\mathcal {L}_2$. We renumber all points $ \mathcal {L}_2 : $
	\begin{equation}\label{v:c} 
	\tau^0=(0,0) ,
	\tau^1=(1,0) , \tau^2=(2,0) ,...,\tau^{L_1}=(L_1,0)\equiv (0,1)
	,\tau^{L}=(0,0)\equiv \tau^0 .
	\end{equation}

	This numeration determines the nature cyclic detour at every point (in the positive direction) and local (cyclic) ordering. 
	Let us consider that there is a particle in each site $t=(t_1,t_2)$. The state of particle is defined by spin $\sigma_t  $  , which at every site of lattice $t=(t_1,t_2)$ can take two values: $\sigma_t \in X=\{+1,-1\} $  . 
	Every spin interacts with the eight nearest spins.  The Hamiltonian of generalized two-dimensional Ising model has the form 
		
	\begin{equation}\label{v:d}
	\begin{gathered}
	\mathcal {H} (\sigma )= -\sum_{m=1}^{L_2} \sum_{n=1}^{L_1}(J_1\sigma_n^m\sigma_{n+1}^m+J_2\sigma_n^m\sigma_{n}^{m+1} + J_3\sigma_{n+1}^m\sigma_{n}^{m+1}+J_4\sigma_n^m\sigma_{n+1}^{m+1}+\\
	+J_5\sigma_n^m\sigma_{n}^{m+1}\sigma_{n+1}^{m+1}+ J_6\sigma_n^m\sigma_{n+1}^{m}\sigma_{n}^{m+1}+   
	J_7\sigma_n^m\sigma_{n+1}^{m}\sigma_{n+1}^{m+1}+ \\
	+J_8\sigma_n^{m+1}\sigma_{n+1}^{m}\sigma_{n+1}^{m+1}+   
	J_9\sigma_n^m\sigma_{n+1}^{m}\sigma_{n}^{m+1}\sigma_{n+1}^{m+1}+
	\hat h \sigma_n^m ,
	\end{gathered} 
	\end{equation}
	
	where $ J_i $, $ i=1,...,9 $ are corresponding coefficients of multi-spin interaction. We rewrite the Hamiltonian (\ref{v:d}) in the form 
		\begin{equation}\label{v:d2}
	\begin{gathered}
	\mathcal {H} (\sigma )= -\sum_{i=0}^{L-1} (J_1\sigma_{	\tau^i}\sigma_{	\tau^{i+1}}+J_2\sigma_{	\tau^i}\sigma_{	\tau^{i+L_1}} + J_3\sigma_{	\tau^{i+1}}\sigma_{	\tau^{i+L_1}}+J_4\sigma_{	\tau^{i}}\sigma_{	\tau^{i+L_1+1}}+\\
	+J_5\sigma_{	\tau^{i}}\sigma_{\tau^{i+L_1}}\sigma_{	\tau^{i+L_1+1}}+ J_6\sigma_{\tau^{i}}\sigma_{\tau^{i+1}}\sigma_{\tau^{i+L_1}}+   
	J_7\sigma_{\tau^{i}}\sigma_{\tau^{i+1}}\sigma_{\tau^{i+L_1+1}}+ \\
	+J_8\sigma_{\tau^{i+L_1}}\sigma_{\tau^{i+1}}\sigma_{\tau^{i+L_1+1}}+   
	J_9\sigma_{\tau^{i}}\sigma_{\tau^{i+1}}\sigma_{\tau^{i+L_1}}
\sigma_{\tau^{i+L_1+1}}	+
	\hat h \sigma_{\tau^{i}} .
	\end{gathered} 
	\end{equation}

	 Partition function of model with the Hamiltonian (\ref{v:d2})  can be written in the following form
	
	\begin{equation}\label{v:e7}
	\begin{split}
	Z_{{L}}=Z_{{L_1}{L_2}}=\sum_{\sigma}\exp ( -\mathcal {H}(\sigma )/{(k_B T)}), 
	\end{split} 
	\end{equation}

	where summation perfomed over all spins.
	
	Let us introduce $ K_i={J_i}/{(k_B T)} $, $ i=1,...,9 $ , where $ T $ is temperature, $k_B $ is Boltzmann's constant, 	$ h'=\hat h /{(k_B T)}  $ is the parameter that determines the interaction with a external field with a coefficient $\hat  h $ . Then partition function (\ref{v:d2}) can be represented in the form 
	
		\begin{equation}\label{v:e2}
	\begin{gathered}
	Z_{{L}}=Z_{{L_1}{L_2}}=\sum_{\sigma}\exp ( \sum_{i=0}^{L-1} (K_1\sigma_{	\tau^i}\sigma_{	\tau^{i+1}}+K_2\sigma_{	\tau^i}\sigma_{	\tau^{i+L_1}} + K_3\sigma_{	\tau^{i+1}}\sigma_{	\tau^{i+L_1}}+\\ +K_4\sigma_{	\tau^{i}}\sigma_{	\tau^{i+L_1+1}}+
	K_5\sigma_{	\tau^{i}}\sigma_{\tau^{i+L_1}}\sigma_{	\tau^{i+L_1+1}}+ K_6\sigma_{\tau^{i}}\sigma_{\tau^{i+1}}\sigma_{\tau^{i+L_1}}+ \\  
	+K_7\sigma_{\tau^{i}}\sigma_{\tau^{i+1}}\sigma_{\tau^{i+L_1+1}}+ 
	K_8\sigma_{\tau^{i+L_1}}\sigma_{\tau^{i+1}}\sigma_{\tau^{i+L_1+1}}+\\ 
	+K_9\sigma_{\tau^{i}}\sigma_{\tau^{i+1}}\sigma_{\tau^{i+L_1}}
	\sigma_{\tau^{i+L_1+1}}	+ h' \sigma_{\tau^{i}}). 
	\end{gathered} 
	\end{equation}

	For using model we write the elementary transfer matrix $ \Theta=\Theta_{p,q} $ of size $ 2^{L_1+1}  \times 2^{L_1+1}$ in the same way, as in \cite{Khrapov3}, \cite{Khrapov5}.  Nonzero elements of the transfer matrix    $ \Theta=\Theta_{p,q} $ are specified by all sorts of pairs of sets of spins
	 $ \{(\sigma_{\tau^0},\sigma_{\tau^{1}},...,\sigma_{\tau^{L_1}}),$ 
	$(\sigma_{\tau^1},\sigma_{\tau^{2}},...,\sigma_{\tau^{L_1+1}})  \} $:
		
	\begin{equation}\label{v:e3}
	\begin{gathered}
	\Theta_{ \{ 
	(\sigma_{\tau^0},\sigma_{\tau^{1}},...,\sigma_{\tau^{L_1}}),
	(\sigma_{\tau^1},\sigma_{\tau^{2}},...,\sigma_{\tau^{L_1+1}})  \}}=
\exp (K_1\sigma_{\tau^0}\sigma_{\tau^1}+K_2\sigma_{\tau^0}\sigma_{\tau^{L_1}} +\\ +K_3\sigma_{\tau^1}\sigma_{\tau^{L_1}}+K_4\sigma_{\tau^1}\sigma_{\tau^{L_1+1}}+
K_5\sigma_{\tau^1}\sigma_{\tau^{L_1}}\sigma_{\tau^{L_1+1}}+ K_6\sigma_{\tau^0}\sigma_{\tau^{1}}\sigma_{\tau^{L_1}}+\\   
+K_7\sigma_{\tau^0}\sigma_{\tau^{1}}\sigma_{\tau^{L_1+1}}+
K_8\sigma_{\tau^{L_1}}\sigma_{\tau^{1}}\sigma_{\tau^{L_1+1}}+  
K_9\sigma_{\tau^0}\sigma_{\tau^{1}}\sigma_{\tau^{L_1}}\sigma_{\tau^{L_1+1}}+\\
+h' \sigma_{\tau^0})=\Theta_{p,q} , 
	\end{gathered} 
	\end{equation}

	 wherein
		\begin{equation}\label{v:e32}
	\begin{split}
	p= \sum_{k=0}^{L_1}{((1-\sigma_{\tau^k})/2)2^k}, p=0,1,\ldots,2^{L_1+1}-1,	
	\end{split} 
	\end{equation}
	
		\begin{equation}\label{v:e4}
	\begin{split}
	q= \sum_{k=0}^{L_1}{((1-\sigma_{\tau^{1+k}})/2)2^k}, q=0,1,\ldots,2^{L_1+1}-1.	
	\end{split} 
	\end{equation}

	Then 
	
		\begin{equation}\label{v:e6}
	\begin{gathered}
		Z_L=Z_{{L_1}{L_2}}=\sum_{ \{\sigma_{\tau^0},\sigma_{\tau^{1}},\ldots,\sigma_{\tau^{L-1}}  \}}
	\Theta_{\{ (\sigma_{\tau^0},\sigma_{\tau^{1}},...,\sigma_{\tau^{L_1}}),
			(\sigma_{\tau^1},\sigma_{\tau^{2}},...,\sigma_{\tau^{L_1+1}})  \} }\\
\Theta_{\{ (\sigma_{\tau^1},\sigma_{\tau^{2}},...,\sigma_{\tau^{L_1+1}}),
			(\sigma_{\tau^2},\sigma_{\tau^{3}},...,\sigma_{\tau^{L_1+2}})  \} }\ldots \\
	\Theta_{\{ (\sigma_{\tau^{L-1}},\sigma_{\tau^{0}},...,\sigma_{\tau^{L_1-1}}),
		(\sigma_{\tau^0},\sigma_{\tau^{1}},...,\sigma_{\tau^{L_1}})  \} } 
	=
	Tr({	\Theta}^L). 
	\end{gathered} 	
	\end{equation}
	
		Now the free energy at one site of lattice $ f $ in the thermodynamic limit can be written in the following form (for example, \cite{baxter2016}) :
	\begin{equation}\label{v:e}
	\begin{split}
	f(T,	 h')=-kT \lim\limits_{L\to {\infty}} {\ln(\lambda_{\max}(L,T,	h'))} , 
	\end{split} 
	\end{equation}
	where $\lambda_{\max}$ is the largest transfer matrix eigenvalue  $ \Theta=\Theta_{p,q}. $
	
	By the Perron–Frobenius theorem \cite{Perron} the only one largest eigenvalue $\lambda_{\max}$ of transfer matrix $ \Theta=\Theta_{p,q} $ will correspond to a matrix with possitive elements (all matrix elements $ \Theta^{L_1+2} $ will strictly be greater than zero, the structure filled with nonzero elements becomes clear already for $ \Theta^{2} $ . Actually, at first the Perron–Frobenius theorem is used for matrix $ \Theta^{L_1+2} $).
	Let us assume
	
	\begin{equation}\label{v:f}  
	\begin{gathered} 
	G(\sigma_{\tau^0},\sigma_{\tau^1},\sigma_{\tau^{L_1}},\sigma_{\tau^{L_1+1}},
	K_1,K_2,...,K_9,h')=\\
	\exp (K_1\sigma_{\tau^0}\sigma_{\tau^1}+K_2\sigma_{\tau^0}\sigma_{\tau^{L_1}} + K_3\sigma_{\tau^1}\sigma_{\tau^{L_1}}+
	K_4\sigma_{\tau^1}\sigma_{\tau^{L_1+1}}+\\
	+K_5\sigma_{\tau^1}\sigma_{\tau^{L_1}}\sigma_{\tau^{L_1+1}}+ K_6\sigma_{\tau^0}\sigma_{\tau^{1}}\sigma_{\tau^{L_1}}+   
	K_7\sigma_{\tau^0}\sigma_{\tau^{1}}\sigma_{\tau^{L_1+1}}+ \\
	+K_8\sigma_{\tau^{L_1}}\sigma_{\tau^{1}}\sigma_{\tau^{L_1+1}}+   
	K_9\sigma_{\tau^0}\sigma_{\tau^{1}}\sigma_{\tau^{L_1}}\sigma_{\tau^{L_1+1}}+
	h' \sigma_{\tau^0}).  
	\end{gathered} 
	\end{equation}
	
	Let us define the values $ a_{ij} $  ,$ j=0,1,...,7 $  , $ i=0,1 $:

	\begin{equation}\label{v:g}  
	\begin{split} 
	a_{i0}=G(+1,+1,+1,1-2i,K_1,K_2,...,K_9,h'),\\
	a_{i1}=G(-1,+1,+1,1-2i,K_1,K_2,...,K_9,h'),\\  
	a_{i2}=G(+1,-1,+1,1-2i,K_1,K_2,...,K_9,h'),\\
	a_{i3}=G(-1,-1,+1,1-2i,K_1,K_2,...,K_9,h'),\\  
	a_{i4}=G(+1,+1,-1,1-2i,K_1,K_2,...,K_9,h'),\\
	a_{i5}=G(-1,+1,-1,1-2i,K_1,K_2,...,K_9,h'),\\  
	a_{i6}=G(+1,-1,-1,1-2i,K_1,K_2,...,K_9,h'),\\
	a_{i7}=G(-1,-1,-1,1-2i,K_1,K_2,...,K_9,h').\\ 
	\end{split} 
	\end{equation}
 Then nonzero elements of matrix $ \Theta $ can be written in the form:
 	\begin{equation}\label{v:g}  
 \begin{gathered} 
 \Theta_{2r+l+4k+s2^{L_1},r+2k+s2^{L_1-1}+i2^{L_1}}=a_{i,2r+l+4s}, \\
 r=0,1,  l=0,1,s=0,1,i=0,1, k=0,1,\ldots,2^{L_1-2}-1.
 \end{gathered} 
 \end{equation}

	\section{Finding the largest eigenvalue and eigenvector of the transfer matrix  }\label{max_val}
	
	The eigenvector of the transfer matrix $ \Theta=\Theta_{p,q} $, corresponding to
	the largest eigenvalue $ F $, we represent in the following form: \\
	
	\begin{enumerate}
		\item Putting $ L_1=2 $, we have \\
		\begin{equation}\label{v:aa}  
		\overrightarrow{x_2}=(1,b_2,c,cb_2;cb_3,cb_4,c^2b_3,c^2b_4)^T 
		 .
		\end{equation}	
		\item 	Let us assume 	$\overrightarrow{x}_{n+1}=(\overrightarrow{x}_{n}^1,
		\overrightarrow{x}_{n}^2)^T $  . Hence 
		\begin{equation}\label{v:bb}  
		\overrightarrow{x}_{n+2}=(\overrightarrow{x}_{n+1}^1,
		\overrightarrow{x}_{n+1}^2)^T ,
		\end{equation}
		where \\
		\begin{equation}\label{v:cc}  
		\overrightarrow{x}_{n+1}^1=(\overrightarrow{x}_{n}^1,c
		\overrightarrow{x}_{n}^1)^T   ,   \overrightarrow{x}_{n+1}^2=(\overrightarrow{x}_{n}^2,c
		\overrightarrow{x}_{n}^2)^T 
		.
		\end{equation}	          
	\end{enumerate}
	Then by the Perron–Frobenius theorem \cite{Perron} this eigenvector with all positive elements will correspond to the single maximal eigenvalue $F$ of the transter matrix $ \Theta=\Theta_{p,q} $. Let us denote 
	\begin{equation}\label{v:dd}  
	\begin{split} 
	R_i=\exp(K_i), i=1,...,9, H=\exp(h')
	\end{split}
	. 
	\end{equation}
	From the form of the elementary transfer matrix  $ \Theta $ (fig.1) and the form of the eigenvector (9) (1.1-1.2), we have at  
	
	\begin{equation}\label{v:ee}  
	R_i>0, i=1,2,\ldots,9, H>0,F>0,b_2>0,b_3>0,b_4>0,c>0
	.
	\end{equation}
	
	the following system of equations 
	\begin{equation}\label{v:ff}  
	\begin{split} 
	\left\{  
	\begin{array}{rcl}  
	F &= & a_{00}+c b_3 a_{10}, \\  
	b_2 F &= & a_{01}+cb_3a_{11}, \\ 
	cF &= & b_2 a_{02}+cb_4 a_{12}, \\ 
	c b_2 F &= & b_2 a_{03}+c b_4 a_{13}, \\ 
	c b_3 F &= & c a_{04}+c^2 b_3 a_{14}, \\
	c b_4 F &= & c a_{05}+c^2 b_3 a_{15}, \\
	c^2 b_3 F &= & c b_2 a_{06}+c^2 b_4 a_{16}, \\
	c^2 b_4 F &= & c b_2 a_{07}+c^2 b_4 a_{17}. \\
	\end{array}   
	\right.
	\end{split} 
	\end{equation}
	
	These eight equations (\ref{v:ff}) will repeat, perhaps, they will just be multiplied by $ c^k $  for some $ k=0,1,2,\ldots  $  .
	Then , using the form of the Hamiltonian (\ref{v:d}), after dividing by $ c^k>0 $ and changing variable (\ref{v:dd}), we have

	\begin{equation}\label{v:gg}  
	\begin{split} 
	\left\{  
	\begin{array}{rcl}  
	F &= & (b_3 c H R_1 R_2 R_3 R_6)/(R_4 R_5 R_7 R_8 R_9) +
	H R_1 R_2 R_3 R_4 R_5 R_6 R_7 R_8 R_9, \; \; \; \; \;\; \;\; \;\; (1)  \\  
	b_2 F &= & (R_3 R_8)/(H R_1 R_2 R_4 R_5 R_6 R_7 R_9) + 
	(b_3 c R_3 R_4 R_5 R_7 R_9)/(H R_1 R_2 R_6 R_8), \; \; \; (2) \\ 
	cF &= & (b_2 H R_2 R_4 R_5)/(R_1 R_3 R_6 R_7 R_8 R_9) + 
	(b_4 c H R_2 R_7 R_8 R_9)/(R_1 R_3 R_4 R_5 R_6),    (3) \\ 
	c b_2 F &= & (b_4 c R_1 R_4 R_5 R_6 R_8)/(H R_2 R_3 R_7 R_9) + (b_2 R_1 R_6 R_7 R_9)/(H R_2 R_3 R_4 R_5 R_8),     (4) \\ 
	b_3 F &= & (H R_1 R_4 R_7)/(R_2 R_3 R_5 R_6 R_8 R_9) + (
	b3 c H R_1 R_5 R_8 R_9)/(R_2 R_3 R_4 R_6 R_7),  \; \; \; (5) \\
	b_4 F &= & (b3 c R_2 R_4 R_6 R_7 R_8)/(H R_1 R_3 R_5 R_9) + (
	R_2 R_5 R_6 R_9)/(H R_1 R_3 R_4 R_7 R_8),     \; \; \; (6) \\
	c b_3 F &= & (b4 c H R_3 R_5 R_6 R_7)/(R_1 R_2 R_4 R_8 R_9) + (
	b2 H R_3 R_4 R_6 R_8 R_9)/(R_1 R_2 R_5 R_7),     (7) \\
	c b_4 F &= & (b2 R_1 R_2 R_3 R_5 R_7 R_8)/(H R_4 R_6 R_9) + (
	b4 c R_1 R_2 R_3 R_4 R_9)/(H R_5 R_6 R_7 R_8).     (8) 
	\end{array}   
	\right.  
	\end{split} 
	\end{equation}
	
		\begin{remark}\label{v:rem1} 
		The system of equations will remain the same (\ref{v:ff})-(\ref{v:gg}) , if in the formula of the Hamiltonian (\ref{v:d2}) we change $ \sigma_{\tau^{i+L_1}} $ and $ \sigma_{\tau^{i+L_1+1}}  $ to $ \sigma_{\tau^{i+L_1+m}} $ and $ \sigma_{\tau^{i+L_1+m+1}}  $ respectively with random integer value $ m>0 $. The  transfer matrix size will only change. This result helps to extend the class of models which disorder solutions are found for. 		
	\end{remark}
	The solution of the system (\ref{v:gg}) is in the section \ref{solution_gg}. Below we write the resulting formulas of solution of the system (\ref{v:gg}), dependeing on seven parameters$ K_1, K_3, K_4,$
	$K_5, K_6, K_7, K_8 $, or, respectively,  $ R_1, R_3, R_4, R_5, R_6, R_7, R_8 $ in order of calculation.
	
	Let us introduce $ u_i=R_i^2,s_i=R_i^4,  i=1,\ldots,9. $ , $ h=H^2 $ and  $ x_1, x_2 $ are the roots of the following quadratic equation 
	\begin{equation}\label{v:hh} 
	A x^2 + B x+ C=0,
	\end{equation}
	where 
	
	\begin{equation}\label{v:ii} 
	\begin{array}{rcl}  
	A =s_1 s_5 s_6 s_7 s_8 - s_3 s_4 s_5 s_6 s_7 s_8 - 
	s_1^2 s_3 s_4 s_5 s_6 s_7 s_8 + 
	s_1 (s_3^2) (s_4^2) s_5 s_6 s_7 s_8, 
	\end{array}  
	\end{equation}

	\begin{equation} \label{v:jj}
	\begin{gathered} 
	B=-s_1 s_3 s_4^2 s_5 s_7 + s_1^2 s_3 s_4 s_6 s_7 + s_3 s_4 s_5^2 s_6 s_7 - 
	s_1 s_3 s_5 s_6^2 s_7 +\\
	+s_1 s_3 s_4 s_5 s_8 - s_1 s_3^2 s_4 s_6 s_8 - 
	s_1 s_4 s_5^2 s_6 s_8 + s_1 s_3 s_4 s_5 s_6^2 s_8 +\\
	+s_1 s_3 s_4 s_5 s_7^2 s_8 - 
	s_1 s_4 s_6 s_7^2 s_8 - s_1 s_3^2 s_4 s_5^2 s_6 s_7^2 s_8 + 
	s_1 s_3 s_4 s_5 s_6^2 s_7^2 s_8 -\\- s_1 s_3 s_5 s_7 s_8^2 + 
	s_3 s_4 s_6 s_7 s_8^2 + 
	s_1^2 s_3 s_4 s_5^2 s_6 s_7 s_8^2 - s_1 s_3 s_4^2 s_5 s_6^2 s_7 s_8^2,
	\end{gathered} 
	\end{equation} 
	
	\begin{multline} \label{v:kk}
	C=s_1 s_3^2 s_5 s_6 s_7 s_8 - s_3 s_4 s_5 s_6 s_7 s_8 - 
	s_1^2 s_3 s_4 s_5 s_6 s_7 s_8 + 
	s_1 s_4^2 s_5 s_6 s_7 s_8.
	\end{multline} 
	
	Hence $ s_9=x_1 $ or $ s_9=x_2 $  ($ s_9>0 $).
	
	\begin{equation} \label{v:ll}
	\begin{gathered}
	h = (u_5 u_6 (u_3^2 u_4^2 u_5^2 u_7^2 - 
	u_1^2 u_3^2 u_6^2 u_7^2 - u_1^2 u_4^2 u_5^2 u_8^2 + 
	u_1^2 u_3^2 u_4^2 u_6^2 u_8^2 + \\
	+u_1^2 u_7^2 u_8^2 u_9^2 - 
	u_3^2 u_4^2 u_7^2 u_8^2 u_9^2) 
	(u_3^2 u_4^2 u_5^2 u_7^2 u_8^2 - u_1^2 u_4^4 u_5^2 u_7^2 u_8^2 - \\
	-u_3^2 u_4^2 u_5^4 u_7^2 u_9^2 + 
	u_1^2 u_3^2 u_5^2 u_6^2 u_7^2 u_9^2 + 
	u_1^2 u_4^2 u_5^4 u_8^2 u_9^2 - 
	u_1^2 u_3^2 u_4^2 u_5^2 u_6^2 u_8^2 u_9^2 + \\
	+u_1^2 u_4^2 u_7^4 u_8^2 u_9^2 
	-u_1^2 u_3^2 u_4^2 u_5^2 u_6^2 u_7^4 u_8^2 u_9^2 - 
	u_3^2 u_4^2 u_7^2 u_8^4 u_9^2 + 
	u_1^2 u_3^2 u_4^4 u_5^2 u_6^2 u_7^2 u_8^4 u_9^2 -\\ 
	-u_1^2 u_5^2 u_7^2 u_8^2 u_9^4 + 
	u_3^2 u_4^2 u_5^2 u_7^2 u_8^2 u_9^4))/ 
	(u_1^2 u_3^2 u_4^2 u_7 u_8 (u_4^2 u_5^2 u_7^2 - \\
	-u_1^2 u_6^2 u_7^2 -
	u_5^2 u_8^2 + u_3^2 u_6^2 u_8^2 + 
	u_1^2 u_5^2 u_6^2 u_7^2 u_8^2 u_9^2 - 
	u_3^2 u_4^2 u_5^2 u_6^2 u_7^2 u_8^2 u_9^2)\\ 
	(u_3^2 u_5^2 - u_1^2 u_4^2 u_5^2 + u_1^2 u_7^2 u_9^2 - 
	u_3^2 u_4^2 u_5^4 u_7^2 u_9^2 - \\
	-u_3^2 u_8^2 u_9^2 + 
	u_1^2 u_4^2 u_5^4 u_8^2 u_9^2 - 
	u_1^2 u_5^2 u_7^2 u_8^2 u_9^4 + 
	u_3^2 u_4^2 u_5^2 u_7^2 u_8^2 u_9^4)),  
	\end{gathered}
	\end{equation} 
	
	\begin{equation} \label{v:mm}
	\begin{gathered}
	c^2 = (u_6 u_8 (u_3^2 u_5^2 - u_1^2 u_4^2 u_5^2 + u_1^2 u_7^2 u_9^2 - 
	+u_1^2 u_3^2 u_5^2 u_6^2 u_7^2 u_9^2 -\\
	-u_3^2 u_8^2 u_9^2 + 
	u_1^2 u_3^2 u_4^2 u_5^2 u_6^2 u_8^2 u_9^2) (-u_4^4 u_5^2 u_7^2 + 
	u_3^2 u_6^2 u_7^2 + u_4^2 u_5^2 u_8^2 -\\- u_3^2 u_4^2 u_6^2 u_8^2 + 
	u_4^2 u_7^4 u_9^2 - u_3^2 u_4^2 u_5^2 u_6^2 u_7^4 u_9^2 - 
	u_7^2 u_8^2 u_9^2 +\\ 
	+u_3^2 u_4^4 u_5^2 u_6^2 u_7^2 u_8^2 u_9^2))/
	(h u_3^2 u_5 u_7 (-u_3^2 u_6^2 +
	u_1^2 u_4^2 u_6^2 - u_4^2 u_7^2 u_9^2 + \\
	+u_3^2 u_4^2 u_5^2 u_6^2 u_7^2 u_9^2 + u_8^2 u_9^2 - 
	u_1^2 u_4^2 u_5^2 u_6^2 u_8^2 u_9^2) (u_4^2 u_5^2 u_7^2 - 
	-u_1^2 u_6^2 u_7^2 -\\- u_5^2 u_8^2 + u_1^2 u_4^2 u_6^2 u_8^2 - 
	u_4^2 u_7^2 u_8^2 u_9^2 + u_1^2 u_5^2 u_6^2 u_7^2 u_8^2 u_9^2 + 
	u_8^4 u_9^2 -\\- u_1^2 u_4^2 u_5^2 u_6^2 u_8^4 u_9^2)).
	\end{gathered}
	\end{equation} 
	
	Then
	
	\begin{equation} \label{v:pp2}
	u_2 = c g_2  ,
	\end{equation} 
	where
	
	\begin{equation} \label{v:nn}
	\begin{gathered}
	g_2 = -(( u_3 u_5^2 u_9^2 (u_3^2 u_4^2 u_5^2 u_7^2 - 
	u_1^2 u_3^2 u_6^2 u_7^2 - u_1^2 u_4^2 u_5^2 u_8^2 + 
	u_1^2 u_3^2 u_4^2 u_6^2 u_8^2 + \\
	+u_1^2 u_7^2 u_8^2 u_9^2 - 
	u_3^2 u_4^2 u_7^2 u_8^2 u_9^2) (u_4^2 u_5^2 u_7^2 - 
	u_1^2 u_6^2 u_7^2 - u_5^2 u_8^2 + 
	u_1^2 u_4^2 u_6^2 u_8^2 -\\- u_4^2 u_7^2 u_8^2 u_9^2 
	+u_1^2 u_5^2 u_6^2 u_7^2 u_8^2 u_9^2 + u_8^4 u_9^2 - 
	u_1^2 u_4^2 u_5^2 u_6^2 u_8^4 u_9^2))/
	(u_4 u_6 u_7 u_8^2 (u_3^2 u_5^2 -\\- u_1^2 u_4^2 u_5^2 + 
	+u_1^2 u_7^2 u_9^2 - 
	u_1^2 u_3^2 u_5^2 u_6^2 u_7^2 u_9^2 - 
	u_3^2 u_8^2 u_9^2 +  
	u_1^2 u_3^2 u_4^2 u_5^2 u_6^2 u_8^2 u_9^2)\\ 
	(-u_3^2 u_5^2 + 
	u_1^2 u_4^2 u_5^2 -u_1^2 u_7^2 u_9^2 + 
	u_3^2 u_4^2 u_5^4 u_7^2 u_9^2 + u_3^2 u_8^2 u_9^2 - 
	u_1^2 u_4^2 u_5^4 u_8^2 u_9^2 +\\+ 
	u_1^2 u_5^2 u_7^2 u_8^2 u_9^4 - 
	u_3^2 u_4^2 u_5^2 u_7^2 u_8^2 u_9^4))),       
	\end{gathered}
	\end{equation} 
	
	\begin{equation} \label{v:rr}
	\begin{gathered}
	b_2 = (c R_1^2 R_3^2 R_7^2 R_8^2 (R_3^4 R_6^4 - 
	R_1^4 R_4^4 R_6^4 + R_4^4 R_7^4 R_9^4 -\\- 
	R_3^4 R_4^4 R_5^4 R_6^4 R_7^4 R_9^4 - R_8^4 R_9^4 + 
	R_1^4 R_4^4 R_5^4 R_6^4 R_8^4 R_9^4))/\\ 
	(R_6^2 R_9^2 (-R_3^4 R_4^4 R_5^4 R_7^4 + R_1^4 R_3^4 R_6^4 R_7^4 + 
	R_1^4 R_4^4 R_5^4 R_8^4 - 
	R_1^4 R_3^4 R_4^4 R_6^4 R_8^4 -\\- 
	R_1^4 R_7^4 R_8^4 R_9^4 + 
	R_3^4 R_4^4 R_7^4 R_8^4 R_9^4)),   
	\end{gathered}
	\end{equation}

	\begin{equation} \label{v:tt}
	\begin{gathered}
	b_3 = (R_4^2 R_7^2 (R_3^4 R_5^4 R_8^2 - 
	R_1^4 R_4^4 R_5^4 R_8^2 + R_1^4 R_7^4 R_8^2 R_9^4 -\\- 
	R_1^4 R_3^4 R_5^4 R_6^4 R_7^4 R_8^2 R_9^4 - 
	R_3^4 R_8^6 R_9^4 +
	R_1^4 R_3^4 R_4^4 R_5^4 R_6^4 R_8^6 R_9^4))/\\ 
	(c R_5^2 R_9^2 (-R_3^4 R_4^4 R_5^4 R_7^4 + R_1^4 R_3^4 R_6^4 R_7^4 + 
	R_1^4 R_4^4 R_5^4 R_8^4 - \\
	-R_1^4 R_3^4 R_4^4 R_6^4 R_8^4 - 
	R_1^4 R_7^4 R_8^4 R_9^4 + 
	R_3^4 R_4^4 R_7^4 R_8^4 R_9^4)),   
	\end{gathered}
	\end{equation} 
	
	\begin{equation} \label{v:uu}
	\begin{gathered}
	b_4 = -((R_4^2 (-R_1^2 R_3^2 R_4^4 R_5^4 R_7^4 + 
	R_1^6 R_3^2 R_6^4 R_7^4 + R_1^2 R_3^2 R_5^4 R_8^4 - 
	R_1^2 R_3^6 R_6^4 R_8^4 - \\
	-R_1^6 R_3^2 R_5^4 R_6^4 R_7^4 R_8^4 R_9^4 + 
	R_1^2 R_3^6 R_4^4 R_5^4 R_6^4 R_7^4 R_8^4 R_9^4))/ \\
	(R_5^2 R_6^2 (R_3^4 R_4^4 R_5^4 R_7^4 - 
	R_1^4 R_3^4 R_6^4 R_7^4 - R_1^4 R_4^4 R_5^4 R_8^4 + \\
	+R_1^4 R_3^4 R_4^4 R_6^4 R_8^4 + 
	R_1^4 R_7^4 R_8^4 R_9^4 - 
	R_3^4 R_4^4 R_7^4 R_8^4 R_9^4))),
	\end{gathered}
	\end{equation}

	\begin{multline} \label{v:vv}
	F = (b3 c H R_1 R_2 R_3 R_6)/(R_4 R_5 R_7 R_8 R_9) + 
	H R_1 R_2 R_3 R_4 R_5 R_6 R_7 R_8 R_9.      
	\end{multline}
	
	Free energy
	
		\begin{equation}\label{v:e8}
	\begin{split}
	-f/(kT)=  \ln(F).
	\end{split} 
	\end{equation}
	
	All formulas are considered on the set of values of variables for which the denominators of the expressions do not vanish.
	
	Example. Let us assume
	
	\begin{equation}\label{v:xx}  
	\begin{split} 
	\begin{array}{rcl}  
	R_1 &= &  1.09331330715597974506367350323500512418,\\
	R_3 &= & 1.79629536359563563762214416192501514895,\\
	R_4 &= & 1.99999999999999995356553575708364740289,\\
	R_5 &= & 0.5,1\\
	R_6 &= & 0.9246508996595731,\\
	R_7 &= & 8.01,\\
	R_8 &= & 1.8967896441086978.
	\end{array} 
	\end{split} 
	\end{equation}

	Using (\ref{v:hh})-(\ref{v:vv}), we find other parameters of disorder solution (in this example we take the following solution $ x=(-B+\sqrt{B^2-4 A C})/(2 A) $ of  (\ref{v:hh}))
	
	\begin{equation}\label{v:yy}  
	\begin{split} 
	\begin{array}{rcl}  
	R_9 &= & 0.925496448962883,\\
	H &= & 0.6752085982927946,\\
	c &= & 0.7545930106949338,\\
	R_2 &= &  0.501744073244163, \\
	b_2 &= & 3.331776110427455,\\
	b_3 &= & 1.792177024209216,\\
	b_4 &= & 3.4301635142941924,\\
	F &= & 8.881632268366628,\\
	-f/(kT) &= & 2.183985354117605.
	\end{array}  
	\end{split} 
	\end{equation}
	  To simplify the process of check let us write all values of variables in the partition function
	  (\ref{v:e2})

	  	\begin{equation}\label{v:yy2}  
	  \begin{split} 	 
	  \begin{array}{rcl}  
	 K_1&= &          0.0892128169345629551, \\
	 K_2&= &         -0.6896651035445359179, \\
	 K_3&= &          0.5857264127035543000, \\
	 K_4&= &          0.6931471805599452862, \\
	 K_5&= &         -0.6733445532637656328, \\
	 K_6&= &         -0.0783390184786288968, \\
	 K_7&= &          2.0806907610802678477, \\
	 K_8&= &          0.6401627960898240710, \\
	 K_9&= &         -0.0774249838609992330, \\
	 h' &=  &        -0.3927336013430913764. 
	  \end{array}  	 
	  \end{split} 
	  \end{equation}

	We write the nonzero values of matrix coefficients $ \Theta=\Theta_{p,q} $ in the matrix form $ a $ 
	
	$$  a= 
	\left(\begin{array}{cccc}  
	a_{00} & a_{10}  \\  
	a_{01} & a_{11}  \\  
	a_{02} & a_{12} \\  
	a_{03} & a_{13}  \\ 
	a_{04} & a_{14}  \\  
	a_{05} & a_{15}  \\  
	a_{06} & a_{16} \\  
	a_{07} & a_{17}  \\ 
	\end{array}\right)  =
	\left(\begin{array}{cccc}  
	8.823624 & 0.042894  \\  
	1.315670 & 20.908524  \\  
	0.013533 & 2.571854 \\  
	6.365205 & 0.433547 \\ 
	15.850509 & 0.049505  \\  
	0.005435 & 22.523530  \\  
	1.757065 & 2.378731 \\  
	6.607026 & 0.377026  \\ 
	\end{array}\right) 
	.
	$$  
	Below there are the eigenvector elements found by numerical power method:

	\begin{equation}\label{v:zz}  
	\begin{split} 
	\begin{array}{rcl}  
	0   &       1.0000000000000000000  \\
	1   &       3.3317761104274707940  \\
	2   &       0.7545930106949368943  \\
	3   &      2.5141349661289202899  \\
	4   &       0.7545930106949367833  \\
	5   &       2.5141349661289216222  \\
	6   &       0.5694106117896466923  \\
	7   &       1.8971486733846267825  \\
	8   &       0.7545930106949366722  \\
	9   &       2.5141349661289211781  \\
	10  &        0.5694106117896465813  \\
	11  &        1.8971486733846278927  \\
	12  &        0.5694106117896465813  \\
	13  &       1.8971486733846274486  \\
	14  &        0.4296732678719935583  \\
	15  &        1.4315751291852043536  \\
	16  &        1.3523642563963256258  \\
	17  &        2.5883774134271795297  \\
	18  &        1.0204846157903186832  \\
	19  &        1.9531715052127802679  \\
	20  &        1.0204846157903184611  \\
	21  &        1.9531715052127798238  \\
	22  &        0.7700505585970789380  \\
	23  &        1.4738495665220672226  \\
	24  &        1.0204846157903184611  \\
	25  &        1.9531715052127811560  \\
	26  &        0.7700505585970787159  \\
	27  &        1.4738495665220667785  \\
	28  &        0.7700505585970792710  \\
	29  &        1.4738495665220661124  \\
	30  &        0.5810747693990848672  \\
	31  &        1.1121565817133087783  \\
	\end{array}   
	\end{split} 
	\end{equation}

	\section{ Solving a system of equations for finding the maximum eigenvalue of a transfer matrix}\label{solution_gg}
		
	The solution scheme for the system of equations (\ref{v:gg}) is following:
	Variable $ F $ from (\ref{v:gg}(1)) we substite in other equations (\ref{v:gg}(2))-(\ref{v:gg}(8)) of the system (\ref{v:gg}). After multiplying denominator and simplifying, we get the system, equivalent to (\ref{v:gg}):

	\begin{equation}\label{v:aaa}  
	\begin{split} 
	\left\{  
	\begin{array}{rcl}  
	Left(1)&= & Right(1),  \; \;\; (1)\\
	c Right(1) &= & Right(3),  \; \;\; (2)\\
	c Right(2) &= & Right(4),   \; \;\; (3)\\
	c Right(5) &= & Right(7),   \; \;\; (4)\\
	c Right(6) &= & Right(8),   \; \;\; (5)\\
	b_2 Right(1) &= & Right(2),   \; \;\; (6)\\
	b_3 Right(1) &= & Right(5),     \; \;\; (7)\\
	b_4 Right(1) &= & Right(6),    \; \; \;  (8)\\
	\end{array}   
	\right.  
	\end{split} 
	\end{equation}
	
	where $ Left(i) $, $ Right(i) $, $ i=1,\dots,9 $ are the left and the right parts of the equation $ i $ of the system (\ref{v:gg}).
	
	After multiplying by positive denominator, simplifying and exception the equation $ (1) $ of the system (\ref{v:aaa}) (in the equation $ (1) $ lets us to get the variable $ F $, not used in the other equations$ (2)-(8) $), we get the following system of equations 
	
	\begin{equation}\label{v:bbb}  
	\begin{split} 
	\left\{  
	\begin{array}{rcl}  
	-b_2 u_4 u_5 + c b_3 c u_1 u_3 u_6 - b_4 c u_7 u_8 u_9 + 
	c u_1 u_3 u_4 u_5 u_6 u_7 u_8 u_9 &= & 0,   \; \;\; (2)\\
	-c u_3 u_8 + b_4 c u_1 u_4 u_5 u_6 u_8 - 
	c b_3 c u_3 u_4 u_5 u_7 u_9 + 
	b_2 u_1 u_6 u_7 u_9 &= & 0,        \; \;\; (3)\\
	c u_1 u_4 u_7 - b_4 c u_3 u_5 u_6 u_7 + 
	c b_3 c u_1 u_5 u_8 u_9 - 
	b_2 u_3 u_4 u_6 u_8 u_9 &= & 0,       \; \;\; (4)\\
	b_2 u_1 u_3 u_5 u_7 u_8 - c b_3 c u_4 u_6 u_7 u_8 + 
	b_4 c u_1 u_3 u_4 u_9 - 
	c u_5 u_6 u_9 &= & 0,   \; \;\; (5)\\
	b_2 b_3 c h u_1 u_2 u_6 - u_8 - b_3 c u_4 u_5 u_7 u_9 + 
	b_2 h u_1 u_2 u_4 u_5 u_6 u_7 u_8 u_9 &= & 0,     \; \;\; (6)\\
	b_3^2 c u_2 u_3 u_6 - u_4 u_7 - b_3 c u_5 u_8 u_9 + 
	b_3 u_2 u_3 u_4 u_5 u_6 u_7 u_8 u_9 &= & 0,       \; \;\; (7)\\
	b_3 b_4 c h u_1 u_3 - b_3 c u_4 u_7 u_8 - u_5 u_9 + 
	b_4 h u_1 u_3 u_4 u_5 u_7 u_8 u_9 &= & 0.    \;\; \;\; (8)
	\end{array}   
	\right.  
	\end{split} 
	\end{equation}
	
	We solve the linear with respect to $ b_2, b_3, b_4 $ subsystem of equations (2,3,4) of the system (\ref{v:bbb}).  We get
	
	\begin{equation}\label{v:ccc}  
	\begin{split} 
	\left\{  
	\begin{array}{lll}  
	b_2~ =  -((
	c u_7 u_8 (-u_1 u_3^3 u_6^2 + u_1^3 u_3 u_4^2 u_6^2 - 
	u_1 u_3 u_4^2 u_7^2 u_9^2 + 
	u_1 u_3^3 u_4^2 u_5^2 u_6^2 u_7^2 u_9^2 + \\
	+u_1 u_3 u_8^2 u_9^2 - 
	-u_1^3 u_3 u_4^2 u_5^2 u_6^2 u_8^2 u_9^2))/(
	u_6 u_9 (-u_3^2 u_4^2 u_5^2 u_7^2 + \\
	+u_1^2 u_3^2 u_6^2 u_7^2 + u_1^2 u_4^2 u_5^2 u_8^2 - 
	u_1^2 u_3^2 u_4^2 u_6^2 u_8^2 - 
	u_1^2 u_7^2 u_8^2 u_9^2 + 
	u_3^2 u_4^2 u_7^2 u_8^2 u_9^2))), \\
	b_3~ =  (-u_3^2 u_4 u_5^2 u_7 u_8 + 
	u_1^2 u_4^3 u_5^2 u_7 u_8 - u_1^2 u_4 u_7^3 u_8 u_9^2 + 
	u_1^2 u_3^2 u_4 u_5^2 u_6^2 u_7^3 u_8 u_9^2 +\\ 
	+u_3^2 u_4 u_7 u_8^3 u_9^2 - 
	u_1^2 u_3^2 u_4^3 u_5^2 u_6^2 u_7 u_8^3 u_9^2)/(
	c u_5 u_9 (u_3^2 u_4^2 u_5^2 u_7^2 - 
	u_1^2 u_3^2 u_6^2 u_7^2 - \\-u_1^2 u_4^2 u_5^2 u_8^2 + 
	u_1^2 u_3^2 u_4^2 u_6^2 u_8^2 + u_1^2 u_7^2 u_8^2 u_9^2 -
	u_3^2 u_4^2 u_7^2 u_8^2 u_9^2)),\\
	b_4~ =   (-u_1 u_3 u_4^3 u_5^2 u_7^2 + 
	u_1^3 u_3 u_4 u_6^2 u_7^2 + u_1 u_3 u_4 u_5^2 u_8^2 - 
	u_1 u_3^3 u_4 u_6^2 u_8^2 - \\
	-u_1^3 u_3 u_4 u_5^2 u_6^2 u_7^2 u_8^2 u_9^2 + 
	u_1 u_3^3 u_4^3 u_5^2 u_6^2 u_7^2 u_8^2 u_9^2)/(
	u_5 u_6 (-u_3^2 u_4^2 u_5^2 u_7^2 + \\
	+u_1^2 u_3^2 u_6^2 u_7^2 + u_1^2 u_4^2 u_5^2 u_8^2 - 
	u_1^2 u_3^2 u_4^2 u_6^2 u_8^2 - u_1^2 u_7^2 u_8^2 u_9^2 +
	u_3^2 u_4^2 u_7^2 u_8^2 u_9^2)).
	\end{array}   
	\right.  
	\end{split} 
	\end{equation}
	
	Substituting $ b_2, b_3, b_4 $ from (\ref{v:ccc}) in the equations (5-8) of the system(\ref{v:bbb}). We remove variables $ b_2, b_3, b_4 $, and after simplifying we have the following system. We write each of four remaining equations separately due to their size. 
	
	1. Equation (\ref{v:bbb})(5):
	
	\begin{multline} \label{v:ddd}
	u_1^2 u_3^4 u_5^2 u_6^2 u_7^2 u_8^2 - 
	u_3^2 u_4^2 u_5^2 u_6^2 u_7^2 u_8^2 - 
	u_1^4 u_3^2 u_4^2 u_5^2 u_6^2 u_7^2 u_8^2 + 
	u_1^2 u_4^4 u_5^2 u_6^2 u_7^2 u_8^2 - \\
	-u_1^2 u_3^2 u_4^4 u_5^2 u_7^2 u_9^2 + 
	u_1^4 u_3^2 u_4^2 u_6^2 u_7^2 u_9^2 + 
	u_3^2 u_4^2 u_5^4 u_6^2 u_7^2 u_9^2 - 
	u_1^2 u_3^2 u_5^2 u_6^4 u_7^2 u_9^2 + \\
	+u_1^2 u_3^2 u_4^2 u_5^2 u_8^2 u_9^2 - 
	u_1^2 u_3^4 u_4^2 u_6^2 u_8^2 u_9^2 - 
	u_1^2 u_4^2 u_5^4 u_6^2 u_8^2 u_9^2 + 
	u_1^2 u_3^2 u_4^2 u_5^2 u_6^4 u_8^2 u_9^2 + \\
	+u_1^2 u_3^2 u_4^2 u_5^2 u_7^4 u_8^2 u_9^2 - 
	u_1^2 u_4^2 u_6^2 u_7^4 u_8^2 u_9^2 - 
	u_1^2 u_3^4 u_4^2 u_5^4 u_6^2 u_7^4 u_8^2 u_9^2 +\\+
	u_1^2 u_3^2 u_4^2 u_5^2 u_6^4 u_7^4 u_8^2 u_9^2 - 
	-u_1^2 u_3^2 u_5^2 u_7^2 u_8^4 u_9^2 + 
	u_3^2 u_4^2 u_6^2 u_7^2 u_8^4 u_9^2 +\\+ 
	u_1^4 u_3^2 u_4^2 u_5^4 u_6^2 u_7^2 u_8^4 u_9^2 - 
	u_1^2 u_3^2 u_4^4 u_5^2 u_6^4 u_7^2 u_8^4 u_9^2 + 
	+u_1^2 u_5^2 u_6^2 u_7^2 u_8^2 u_9^4 - \\-
	u_3^2 u_4^2 u_5^2 u_6^2 u_7^2 u_8^2 u_9^4 - 
	u_1^4 u_3^2 u_4^2 u_5^2 u_6^2 u_7^2 u_8^2 u_9^4 +
	u_1^2 u_3^4 u_4^4 u_5^2 u_6^2 u_7^2 u_8^2 u_9^4 = 0.   
	\end{multline} 
	
	2. Equation (\ref{v:bbb})(6):
	
	\begin{multline} \label{v:eee}
c h u_2 u_3^5 u_4 u_5^2 u_6^2 u_7^2 u_8 - 
2 c h u_1^2 u_2 u_3^3 u_4^3 u_5^2 u_6^2 u_7^2 u_8 + 
c h u_1^4 u_2 u_3 u_4^5 u_5^2 u_6^2 u_7^2 u_8 - \\-
u_3^2 u_4^6 u_5^5 u_7^4 u_9^2 + u_3^4 u_4^2 u_5^3 u_6^2 u_7^4 u_9^2 + 
u_1^2 u_3^2 u_4^4 u_5^3 u_6^2 u_7^4 u_9^2 -\\- u_1^2 u_3^4 u_5 u_6^4 u_7^4 u_9^2 +
c h u_2 u_3^3 u_4^3 u_5^2 u_7^4 u_8 u_9^2 - 
c h u_1^2 u_2 u_3 u_4^5 u_5^2 u_7^4 u_8 u_9^2 +\\+ 
c h u_1^2 u_2 u_3^3 u_4 u_6^2 u_7^4 u_8 u_9^2 - 
c h u_1^4 u_2 u_3 u_4^3 u_6^2 u_7^4 u_8 u_9^2 - 
2 c h u_2 u_3^5 u_4^3 u_5^4 u_6^2 u_7^4 u_8 u_9^2 + \\+
2 c h u_1^2 u_2 u_3^3 u_4^5 u_5^4 u_6^2 u_7^4 u_8 u_9^2 +
u_3^2 u_4^4 u_5^5 u_7^2 u_8^2 u_9^2 + u_1^2 u_4^6 u_5^5 u_7^2 u_8^2 u_9^2 - \\-
2 u_1^2 u_3^2 u_4^2 u_5^3 u_6^2 u_7^2 u_8^2 u_9^2 - 
u_3^4 u_4^4 u_5^3 u_6^2 u_7^2 u_8^2 u_9^2 - \\-
u_1^2 u_3^2 u_4^6 u_5^3 u_6^2 u_7^2 u_8^2 u_9^2 + 
2 u_1^2 u_3^4 u_4^2 u_5 u_6^4 u_7^2 u_8^2 u_9^2 - \\-
c h u_2 u_3^3 u_4 u_5^2 u_7^2 u_8^3 u_9^2 + 
c h u_1^2 u_2 u_3 u_4^3 u_5^2 u_7^2 u_8^3 u_9^2 - \\-
c h u_2 u_3^5 u_4 u_6^2 u_7^2 u_8^3 u_9^2 + 
c h u_1^2 u_2 u_3^3 u_4^3 u_6^2 u_7^2 u_8^3 u_9^2 + \\+
2 c h u_1^2 u_2 u_3^3 u_4^3 u_5^4 u_6^2 u_7^2 u_8^3 u_9^2 - 
2 c h u_1^4 u_2 u_3 u_4^5 u_5^4 u_6^2 u_7^2 u_8^3 u_9^2 - \\-
u_1^2 u_4^4 u_5^5 u_8^4 u_9^2 + 2 u_1^2 u_3^2 u_4^4 u_5^3 u_6^2 u_8^4 u_9^2 - 
u_1^2 u_3^4 u_4^4 u_5 u_6^4 u_8^4 u_9^2 + u_3^2 u_4^4 u_5^3 u_7^6 u_9^4 - \\-
u_1^2 u_3^2 u_4^2 u_5 u_6^2 u_7^6 u_9^4 - u_3^4 u_4^4 u_5^5 u_6^2 u_7^6 u_9^4 + 
u_1^2 u_3^4 u_4^2 u_5^3 u_6^4 u_7^6 u_9^4 + \\+
c h u_1^2 u_2 u_3 u_4^3 u_7^6 u_8 u_9^4 - 
c h u_2 u_3^3 u_4^5 u_5^4 u_7^6 u_8 u_9^4 - \\-
c h u_1^2 u_2 u_3^3 u_4^3 u_5^2 u_6^2 u_7^6 u_8 u_9^4 + 
c h u_2 u_3^5 u_4^5 u_5^6 u_6^2 u_7^6 u_8 u_9^4 - 
u_3^2 u_4^2 u_5^3 u_7^4 u_8^2 u_9^4 -\\- 2 u_1^2 u_4^4 u_5^3 u_7^4 u_8^2 u_9^4 + 
u_3^2 u_4^6 u_5^3 u_7^4 u_8^2 u_9^4 + \\+
2 u_1^2 u_3^2 u_5 u_6^2 u_7^4 u_8^2 u_9^4 - 
u_3^4 u_4^2 u_5 u_6^2 u_7^4 u_8^2 u_9^4 + 
u_1^2 u_3^2 u_4^4 u_5 u_6^2 u_7^4 u_8^2 u_9^4 +\\+ 
u_1^2 u_3^2 u_4^4 u_5^5 u_6^2 u_7^4 u_8^2 u_9^4 + 
u_3^4 u_4^6 u_5^5 u_6^2 u_7^4 u_8^2 u_9^4 - 
2 u_1^2 u_3^4 u_4^4 u_5^3 u_6^4 u_7^4 u_8^2 u_9^4 - \\-
c h u_1^2 u_2 u_3 u_4 u_7^4 u_8^3 u_9^4 - 
c h u_2 u_3^3 u_4^3 u_7^4 u_8^3 u_9^4 + \\+
c h u_2 u_3^3 u_4^3 u_5^4 u_7^4 u_8^3 u_9^4 + 
c h u_1^2 u_2 u_3 u_4^5 u_5^4 u_7^4 u_8^3 u_9^4 - \\-
c h u_1^2 u_2 u_3^3 u_4 u_5^2 u_6^2 u_7^4 u_8^3 u_9^4 + 
2 c h u_1^4 u_2 u_3 u_4^3 u_5^2 u_6^2 u_7^4 u_8^3 u_9^4 + \\+
2 c h u_2 u_3^5 u_4^3 u_5^2 u_6^2 u_7^4 u_8^3 u_9^4 - 
c h u_1^2 u_2 u_3^3 u_4^5 u_5^2 u_6^2 u_7^4 u_8^3 u_9^4 - \\-
2 c h u_1^2 u_2 u_3^3 u_4^5 u_5^6 u_6^2 u_7^4 u_8^3 u_9^4 + 
2 u_1^2 u_4^2 u_5^3 u_7^2 u_8^4 u_9^4 - u_3^2 u_4^4 u_5^3 u_7^2 u_8^4 u_9^4 - \\-
2 u_1^2 u_3^2 u_4^2 u_5 u_6^2 u_7^2 u_8^4 u_9^4 + 
u_3^4 u_4^4 u_5 u_6^2 u_7^2 u_8^4 u_9^4 - \\-
u_1^2 u_3^2 u_4^6 u_5^5 u_6^2 u_7^2 u_8^4 u_9^4 + 
u_1^2 u_3^4 u_4^6 u_5^3 u_6^4 u_7^2 u_8^4 u_9^4 + \\+
c h u_2 u_3^3 u_4 u_7^2 u_8^5 u_9^4 - 
c h u_1^2 u_2 u_3 u_4^3 u_5^4 u_7^2 u_8^5 u_9^4 - \\-
c h u_1^2 u_2 u_3^3 u_4^3 u_5^2 u_6^2 u_7^2 u_8^5 u_9^4 + 
c h u_1^4 u_2 u_3 u_4^5 u_5^6 u_6^2 u_7^2 u_8^5 u_9^4 + \\+
u_1^2 u_4^2 u_5 u_7^6 u_8^2 u_9^6 - u_3^2 u_4^4 u_5 u_7^6 u_8^2 u_9^6 - 
u_1^2 u_3^2 u_4^2 u_5^3 u_6^2 u_7^6 u_8^2 u_9^6 + \\+
u_3^4 u_4^4 u_5^3 u_6^2 u_7^6 u_8^2 u_9^6 - 
c h u_1^2 u_2 u_3 u_4^3 u_5^2 u_7^6 u_8^3 u_9^6 + \\+
c h u_2 u_3^3 u_4^5 u_5^2 u_7^6 u_8^3 u_9^6 + 
c h u_1^2 u_2 u_3^3 u_4^3 u_5^4 u_6^2 u_7^6 u_8^3 u_9^6 -\\ -
c h u_2 u_3^5 u_4^5 u_5^4 u_6^2 u_7^6 u_8^3 u_9^6 - 
u_1^2 u_5 u_7^4 u_8^4 u_9^6 + u_3^2 u_4^2 u_5 u_7^4 u_8^4 u_9^6 + \\+
u_1^2 u_3^2 u_4^4 u_5^3 u_6^2 u_7^4 u_8^4 u_9^6 - 
u_3^4 u_4^6 u_5^3 u_6^2 u_7^4 u_8^4 u_9^6 + \\+
c h u_1^2 u_2 u_3 u_4 u_5^2 u_7^4 u_8^5 u_9^6 - 
c h u_2 u_3^3 u_4^3 u_5^2 u_7^4 u_8^5 u_9^6 - \\-
c h u_1^4 u_2 u_3 u_4^3 u_5^4 u_6^2 u_7^4 u_8^5 u_9^6 + 
c h u_1^2 u_2 u_3^3 u_4^5 u_5^4 u_6^2 u_7^4 u_8^5 u_9^6 = 0. 
	\end{multline}

	3. Equation (\ref{v:bbb})(7):
	
	\begin{multline} \label{v:fff}
	-u_2 u_3^4 u_4 u_5^4 u_6 u_7 u_8^2 + 
	2 u_1^2 u_2 u_3^2 u_4^3 u_5^4 u_6 u_7 u_8^2 - 
	u_1^4 u_2 u_4^5 u_5^4 u_6 u_7 u_8^2 + \\+
	c u_3^3 u_4^4 u_5^6 u_7^4 u_9^2 - 
	2 c u_1^2 u_3^3 u_4^2 u_5^4 u_6^2 u_7^4 u_9^2 + 
	c u_1^4 u_3^3 u_5^2 u_6^4 u_7^4 u_9^2 - \\-
	c u_3^3 u_4^2 u_5^6 u_7^2 u_8^2 u_9^2 - 
	c u_1^2 u_3 u_4^4 u_5^6 u_7^2 u_8^2 u_9^2 + 
	c u_1^2 u_3^3 u_5^4 u_6^2 u_7^2 u_8^2 u_9^2 + \\+
	c u_1^4 u_3 u_4^2 u_5^4 u_6^2 u_7^2 u_8^2 u_9^2 + 
	2 c u_1^2 u_3^3 u_4^4 u_5^4 u_6^2 u_7^2 u_8^2 u_9^2 - 
	2 c u_1^4 u_3^3 u_4^2 u_5^2 u_6^4 u_7^2 u_8^2 u_9^2 - \\-
	2 u_1^2 u_2 u_3^2 u_4 u_5^2 u_6 u_7^3 u_8^2 u_9^2 + 
	2 u_1^4 u_2 u_4^3 u_5^2 u_6 u_7^3 u_8^2 u_9^2 + 
	u_2 u_3^4 u_4^3 u_5^6 u_6 u_7^3 u_8^2 u_9^2 - \\-
	u_1^2 u_2 u_3^2 u_4^5 u_5^6 u_6 u_7^3 u_8^2 u_9^2 + 
	u_1^2 u_2 u_3^4 u_4 u_5^4 u_6^3 u_7^3 u_8^2 u_9^2 - 
	u_1^4 u_2 u_3^2 u_4^3 u_5^4 u_6^3 u_7^3 u_8^2 u_9^2 + \\+
	c u_1^2 u_3 u_4^2 u_5^6 u_8^4 u_9^2 - 
	c u_1^2 u_3^3 u_4^2 u_5^4 u_6^2 u_8^4 u_9^2 - 
	c u_1^4 u_3 u_4^4 u_5^4 u_6^2 u_8^4 u_9^2 + \\+
	c u_1^4 u_3^3 u_4^4 u_5^2 u_6^4 u_8^4 u_9^2 + 
	2 u_2 u_3^4 u_4 u_5^2 u_6 u_7 u_8^4 u_9^2 - 
	2 u_1^2 u_2 u_3^2 u_4^3 u_5^2 u_6 u_7 u_8^4 u_9^2 - \\-
	u_1^2 u_2 u_3^2 u_4^3 u_5^6 u_6 u_7 u_8^4 u_9^2 + 
	u_1^4 u_2 u_4^5 u_5^6 u_6 u_7 u_8^4 u_9^2 - 
	u_1^2 u_2 u_3^4 u_4^3 u_5^4 u_6^3 u_7 u_8^4 u_9^2 + \\+
	u_1^4 u_2 u_3^2 u_4^5 u_5^4 u_6^3 u_7 u_8^4 u_9^2 + 
	c u_1^2 u_3 u_4^2 u_5^4 u_7^4 u_8^2 u_9^4 - 
	2 c u_3^3 u_4^4 u_5^4 u_7^4 u_8^2 u_9^4 - \\-
	c u_1^4 u_3 u_5^2 u_6^2 u_7^4 u_8^2 u_9^4 + 
	2 c u_1^2 u_3^3 u_4^2 u_5^2 u_6^2 u_7^4 u_8^2 u_9^4 + 
	c u_1^2 u_3^3 u_4^2 u_5^6 u_6^2 u_7^4 u_8^2 u_9^4 - 
	c u_1^4 u_3^3 u_5^4 u_6^4 u_7^4 u_8^2 u_9^4 - \\-
	u_1^4 u_2 u_4 u_6 u_7^5 u_8^2 u_9^4 + 
	u_1^2 u_2 u_3^2 u_4^3 u_5^4 u_6 u_7^5 u_8^2 u_9^4 + 
	u_1^4 u_2 u_3^2 u_4 u_5^2 u_6^3 u_7^5 u_8^2 u_9^4 - \\-
	u_1^2 u_2 u_3^4 u_4^3 u_5^6 u_6^3 u_7^5 u_8^2 u_9^4 -
	c u_1^2 u_3 u_5^4 u_7^2 u_8^4 u_9^4 + 
	2 c u_3^3 u_4^2 u_5^4 u_7^2 u_8^4 u_9^4 + \\+
	c u_1^2 u_3 u_4^4 u_5^4 u_7^2 u_8^4 u_9^4 - 
	c u_1^2 u_3^3 u_5^2 u_6^2 u_7^2 u_8^4 u_9^4 + 
	c u_1^4 u_3 u_4^2 u_5^2 u_6^2 u_7^2 u_8^4 u_9^4 - \\-
	2 c u_1^2 u_3^3 u_4^4 u_5^2 u_6^2 u_7^2 u_8^4 u_9^4 - 
	c u_1^4 u_3 u_4^2 u_5^6 u_6^2 u_7^2 u_8^4 u_9^4 - 
	c u_1^2 u_3^3 u_4^4 u_5^6 u_6^2 u_7^2 u_8^4 u_9^4 + \\+
	2 c u_1^4 u_3^3 u_4^2 u_5^4 u_6^4 u_7^2 u_8^4 u_9^4 + 
	2 u_1^2 u_2 u_3^2 u_4 u_6 u_7^3 u_8^4 u_9^4 + 
	u_1^2 u_2 u_3^2 u_4 u_5^4 u_6 u_7^3 u_8^4 u_9^4 - \\-
	2 u_1^4 u_2 u_4^3 u_5^4 u_6 u_7^3 u_8^4 u_9^4 - 
	2 u_2 u_3^4 u_4^3 u_5^4 u_6 u_7^3 u_8^4 u_9^4 + 
	u_1^2 u_2 u_3^2 u_4^5 u_5^4 u_6 u_7^3 u_8^4 u_9^4 - \\-
	u_1^2 u_2 u_3^4 u_4 u_5^2 u_6^3 u_7^3 u_8^4 u_9^4 - 
	u_1^4 u_2 u_3^2 u_4^3 u_5^2 u_6^3 u_7^3 u_8^4 u_9^4 + 
	u_1^4 u_2 u_3^2 u_4^3 u_5^6 u_6^3 u_7^3 u_8^4 u_9^4 + \\+
	u_1^2 u_2 u_3^4 u_4^5 u_5^6 u_6^3 u_7^3 u_8^4 u_9^4 - 
	c u_1^2 u_3 u_4^2 u_5^4 u_8^6 u_9^4 + 
	c u_1^2 u_3^3 u_4^2 u_5^2 u_6^2 u_8^6 u_9^4 + \\+
	c u_1^4 u_3 u_4^4 u_5^6 u_6^2 u_8^6 u_9^4 - 
	c u_1^4 u_3^3 u_4^4 u_5^4 u_6^4 u_8^6 u_9^4 - 
	u_2 u_3^4 u_4 u_6 u_7 u_8^6 u_9^4 + \\+
	u_1^2 u_2 u_3^2 u_4^3 u_5^4 u_6 u_7 u_8^6 u_9^4 + 
	u_1^2 u_2 u_3^4 u_4^3 u_5^2 u_6^3 u_7 u_8^6 u_9^4 - 
	u_1^4 u_2 u_3^2 u_4^5 u_5^6 u_6^3 u_7 u_8^6 u_9^4 - \\-
	c u_1^2 u_3 u_4^2 u_5^2 u_7^4 u_8^4 u_9^6 + 
	c u_3^3 u_4^4 u_5^2 u_7^4 u_8^4 u_9^6 + 
	c u_1^4 u_3 u_5^4 u_6^2 u_7^4 u_8^4 u_9^6 - \\-
	c u_1^2 u_3^3 u_4^2 u_5^4 u_6^2 u_7^4 u_8^4 u_9^6 + 
	u_1^4 u_2 u_4 u_5^2 u_6 u_7^5 u_8^4 u_9^6 - 
	u_1^2 u_2 u_3^2 u_4^3 u_5^2 u_6 u_7^5 u_8^4 u_9^6 - \\-
	u_1^4 u_2 u_3^2 u_4 u_5^4 u_6^3 u_7^5 u_8^4 u_9^6 + 
	u_1^2 u_2 u_3^4 u_4^3 u_5^4 u_6^3 u_7^5 u_8^4 u_9^6 + 
	c u_1^2 u_3 u_5^2 u_7^2 u_8^6 u_9^6 - \\-
	c u_3^3 u_4^2 u_5^2 u_7^2 u_8^6 u_9^6 - 
	c u_1^4 u_3 u_4^2 u_5^4 u_6^2 u_7^2 u_8^6 u_9^6 + 
	c u_1^2 u_3^3 u_4^4 u_5^4 u_6^2 u_7^2 u_8^6 u_9^6 - \\-
	u_1^2 u_2 u_3^2 u_4 u_5^2 u_6 u_7^3 u_8^6 u_9^6 + 
	u_2 u_3^4 u_4^3 u_5^2 u_6 u_7^3 u_8^6 u_9^6 + \\+
	u_1^4 u_2 u_3^2 u_4^3 u_5^4 u_6^3 u_7^3 u_8^6 u_9^6 - 
	u_1^2 u_2 u_3^4 u_4^5 u_5^4 u_6^3 u_7^3 u_8^6 u_9^6 = 0.   
	\end{multline}

	4. Equation (\ref{v:bbb})(8):
	
	\begin{multline} \label{v:ggg}
	h u_1^2 u_3^4 u_4^4 u_5^4 u_7^3 u_8 - 
	h u_1^4 u_3^2 u_4^6 u_5^4 u_7^3 u_8 - 
	h u_1^4 u_3^4 u_4^2 u_5^2 u_6^2 u_7^3 u_8 + 
	h u_1^6 u_3^2 u_4^4 u_5^2 u_6^2 u_7^3 u_8 - \\-
	u_3^4 u_4^4 u_5^5 u_6 u_7^4 u_8^2 + 
	u_1^2 u_3^2 u_4^6 u_5^5 u_6 u_7^4 u_8^2 + 
	u_1^2 u_3^4 u_4^2 u_5^3 u_6^3 u_7^4 u_8^2 - 
	u_1^4 u_3^2 u_4^4 u_5^3 u_6^3 u_7^4 u_8^2 - \\-
	h u_1^2 u_3^4 u_4^2 u_5^4 u_7 u_8^3 + 
	h u_1^4 u_3^2 u_4^4 u_5^4 u_7 u_8^3 + 
	h u_1^2 u_3^6 u_4^2 u_5^2 u_6^2 u_7 u_8^3 - 
	h u_1^4 u_3^4 u_4^4 u_5^2 u_6^2 u_7 u_8^3 + \\+
	u_1^2 u_3^2 u_4^4 u_5^5 u_6 u_7^2 u_8^4 - 
	u_1^4 u_4^6 u_5^5 u_6 u_7^2 u_8^4 - 
	u_1^2 u_3^4 u_4^4 u_5^3 u_6^3 u_7^2 u_8^4 + 
	u_1^4 u_3^2 u_4^6 u_5^3 u_6^3 u_7^2 u_8^4 + \\+
	u_3^4 u_4^4 u_5^7 u_6 u_7^4 u_9^2 - 
	2 u_1^2 u_3^4 u_4^2 u_5^5 u_6^3 u_7^4 u_9^2 + 
	u_1^4 u_3^4 u_5^3 u_6^5 u_7^4 u_9^2 + 
	h u_1^4 u_3^2 u_4^4 u_5^2 u_7^5 u_8 u_9^2 - \\-
	h u_1^2 u_3^4 u_4^6 u_5^6 u_7^5 u_8 u_9^2 - 
	h u_1^6 u_3^2 u_4^2 u_6^2 u_7^5 u_8 u_9^2 + 
	h u_1^4 u_3^4 u_4^4 u_5^4 u_6^2 u_7^5 u_8 u_9^2 - \\-
	2 u_1^2 u_3^2 u_4^4 u_5^7 u_6 u_7^2 u_8^2 u_9^2 + 
	2 u_1^4 u_3^2 u_4^2 u_5^5 u_6^3 u_7^2 u_8^2 u_9^2 + 
	2 u_1^2 u_3^4 u_4^4 u_5^5 u_6^3 u_7^2 u_8^2 u_9^2 - \\-
	2 u_1^4 u_3^4 u_4^2 u_5^3 u_6^5 u_7^2 u_8^2 u_9^2 - 
	u_1^2 u_3^2 u_4^4 u_5^3 u_6 u_7^6 u_8^2 u_9^2 + 
	u_1^4 u_3^2 u_4^2 u_5 u_6^3 u_7^6 u_8^2 u_9^2 + \\+
	u_1^2 u_3^4 u_4^4 u_5^5 u_6^3 u_7^6 u_8^2 u_9^2 - 
	u_1^4 u_3^4 u_4^2 u_5^3 u_6^5 u_7^6 u_8^2 u_9^2 - 
	h u_1^4 u_3^2 u_4^2 u_5^2 u_7^3 u_8^3 u_9^2 - \\-
	h u_1^2 u_3^4 u_4^4 u_5^2 u_7^3 u_8^3 u_9^2 + 
	h u_1^2 u_3^4 u_4^4 u_5^6 u_7^3 u_8^3 u_9^2 + 
	h u_1^4 u_3^2 u_4^6 u_5^6 u_7^3 u_8^3 u_9^2 + \\+
	2 h u_1^4 u_3^4 u_4^2 u_6^2 u_7^3 u_8^3 u_9^2 + 
	h u_1^4 u_3^4 u_4^2 u_5^4 u_6^2 u_7^3 u_8^3 u_9^2 - 
	2 h u_1^6 u_3^2 u_4^4 u_5^4 u_6^2 u_7^3 u_8^3 u_9^2 - \\-
	2 h u_1^2 u_3^6 u_4^4 u_5^4 u_6^2 u_7^3 u_8^3 u_9^2 + 
	h u_1^4 u_3^4 u_4^6 u_5^4 u_6^2 u_7^3 u_8^3 u_9^2 + 
	u_1^4 u_4^4 u_5^7 u_6 u_8^4 u_9^2 - \\
	2 u_1^4 u_3^2 u_4^4 u_5^5 u_6^3 u_8^4 u_9^2 + 
	u_1^4 u_3^4 u_4^4 u_5^3 u_6^5 u_8^4 u_9^2 - 
	u_1^2 u_3^2 u_4^2 u_5^3 u_6 u_7^4 u_8^4 u_9^2 + \\+
	2 u_1^4 u_4^4 u_5^3 u_6 u_7^4 u_8^4 u_9^2 + 
	2 u_3^4 u_4^4 u_5^3 u_6 u_7^4 u_8^4 u_9^2 - 
	u_1^2 u_3^2 u_4^6 u_5^3 u_6 u_7^4 u_8^4 u_9^2 - \\-
	u_1^2 u_3^4 u_4^2 u_5 u_6^3 u_7^4 u_8^4 u_9^2 - 
	u_1^4 u_3^2 u_4^4 u_5 u_6^3 u_7^4 u_8^4 u_9^2 - 
	u_1^4 u_3^2 u_4^4 u_5^5 u_6^3 u_7^4 u_8^4 u_9^2 - \\-
	u_1^2 u_3^4 u_4^6 u_5^5 u_6^3 u_7^4 u_8^4 u_9^2 + 
	2 u_1^4 u_3^4 u_4^4 u_5^3 u_6^5 u_7^4 u_8^4 u_9^2 + 
	h u_1^2 u_3^4 u_4^2 u_5^2 u_7 u_8^5 u_9^2 - \\
	h u_1^4 u_3^2 u_4^4 u_5^6 u_7 u_8^5 u_9^2 - 
	h u_1^2 u_3^6 u_4^2 u_6^2 u_7 u_8^5 u_9^2 + 
	h u_1^4 u_3^4 u_4^4 u_5^4 u_6^2 u_7 u_8^5 u_9^2 - \\-
	u_1^2 u_3^2 u_4^4 u_5^3 u_6 u_7^2 u_8^6 u_9^2 + 
	u_1^2 u_3^4 u_4^4 u_5 u_6^3 u_7^2 u_8^6 u_9^2 + 
	u_1^4 u_3^2 u_4^6 u_5^5 u_6^3 u_7^2 u_8^6 u_9^2 - \\-
	u_1^4 u_3^4 u_4^6 u_5^3 u_6^5 u_7^2 u_8^6 u_9^2 + 
	2 u_1^2 u_3^2 u_4^2 u_5^5 u_6 u_7^4 u_8^2 u_9^4 - 
	2 u_3^4 u_4^4 u_5^5 u_6 u_7^4 u_8^2 u_9^4 - \\-
	2 u_1^4 u_3^2 u_5^3 u_6^3 u_7^4 u_8^2 u_9^4 + 
	2 u_1^2 u_3^4 u_4^2 u_5^3 u_6^3 u_7^4 u_8^2 u_9^4 - 
	h u_1^4 u_3^2 u_4^4 u_5^4 u_7^5 u_8^3 u_9^4 + \\+
	h u_1^2 u_3^4 u_4^6 u_5^4 u_7^5 u_8^3 u_9^4 + 
	2 h u_1^6 u_3^2 u_4^2 u_5^2 u_6^2 u_7^5 u_8^3 u_9^4 - 
	2 h u_1^4 u_3^4 u_4^4 u_5^2 u_6^2 u_7^5 u_8^3 u_9^4 - \\-
	h u_1^4 u_3^4 u_4^4 u_5^6 u_6^2 u_7^5 u_8^3 u_9^4 + 
	h u_1^2 u_3^6 u_4^6 u_5^6 u_6^2 u_7^5 u_8^3 u_9^4 - 
	2 u_1^4 u_4^2 u_5^5 u_6 u_7^2 u_8^4 u_9^4 + \\+
	2 u_1^2 u_3^2 u_4^4 u_5^5 u_6 u_7^2 u_8^4 u_9^4 + 
	2 u_1^4 u_3^2 u_4^2 u_5^3 u_6^3 u_7^2 u_8^4 u_9^4 - 
	2 u_1^2 u_3^4 u_4^4 u_5^3 u_6^3 u_7^2 u_8^4 u_9^4 - \\-
	u_1^4 u_4^2 u_5 u_6 u_7^6 u_8^4 u_9^4 + 
	u_1^2 u_3^2 u_4^4 u_5 u_6 u_7^6 u_8^4 u_9^4 + 
	u_1^4 u_3^2 u_4^2 u_5^3 u_6^3 u_7^6 u_8^4 u_9^4 - \\-
	u_1^2 u_3^4 u_4^4 u_5^3 u_6^3 u_7^6 u_8^4 u_9^4 + 
	h u_1^4 u_3^2 u_4^2 u_5^4 u_7^3 u_8^5 u_9^4 - 
	h u_1^2 u_3^4 u_4^4 u_5^4 u_7^3 u_8^5 u_9^4 - \\-
	2 h u_1^4 u_3^4 u_4^2 u_5^2 u_6^2 u_7^3 u_8^5 u_9^4 + 
	2 h u_1^2 u_3^6 u_4^4 u_5^2 u_6^2 u_7^3 u_8^5 u_9^4 + 
	h u_1^6 u_3^2 u_4^4 u_5^6 u_6^2 u_7^3 u_8^5 u_9^4 - \\-
	h u_1^4 u_3^4 u_4^6 u_5^6 u_6^2 u_7^3 u_8^5 u_9^4 + 
	u_1^2 u_3^2 u_4^2 u_5 u_6 u_7^4 u_8^6 u_9^4 - 
	u_3^4 u_4^4 u_5 u_6 u_7^4 u_8^6 u_9^4 - \\-
	u_1^4 u_3^2 u_4^4 u_5^3 u_6^3 u_7^4 u_8^6 u_9^4 + 
	u_1^2 u_3^4 u_4^6 u_5^3 u_6^3 u_7^4 u_8^6 u_9^4 + 
	u_1^4 u_5^3 u_6 u_7^4 u_8^4 u_9^6 - \\-
	2 u_1^2 u_3^2 u_4^2 u_5^3 u_6 u_7^4 u_8^4 u_9^6 + 
	u_3^4 u_4^4 u_5^3 u_6 u_7^4 u_8^4 u_9^6 - 
	h u_1^6 u_3^2 u_4^2 u_5^4 u_6^2 u_7^5 u_8^5 u_9^6 + \\+
	2 h u_1^4 u_3^4 u_4^4 u_5^4 u_6^2 u_7^5 u_8^5 u_9^6 - 
	h u_1^2 u_3^6 u_4^6 u_5^4 u_6^2 u_7^5 u_8^5 u_9^6 = 0.   
	\end{multline} 
	
	Using (\ref{v:ggg}), we express $ h $ and obtain (\ref{v:ll}).

	Using (\ref{v:fff}), we get (\ref{v:pp2}) and (\ref{v:nn}). 	
	Substituting $ u_2 $ from (\ref{v:pp2})-(\ref{v:nn})  in  (\ref{v:eee}), we have
	$ c^2 $ in (\ref{v:mm}).	
	


\begin{thebibliography}{99}


		\bibitem{Onsager}
		\newblock Onsager L.,
		\newblock Crystal statistics. I. A two-dimensional model with an order–disorder transition,
		\newblock \emph{Physical Review}, \textbf{65(3-4)} ( 1944), 117-149. DOI: 10.1103/Phys-Rev.65.117
		
		\bibitem{Wu1982}
		\newblock Wu F.Y. ,
		\newblock The Potts model.  
		\newblock \emph{Reviews of Modern Physics.}, \textbf{54(1)}  (1982), 235-268. DOI: 10.1103/RevModPhys.54.235
		
		\bibitem{baxter2016}
		\newblock Baxter, R.J.,
		\newblock Exactly Solved Models in Statistical Mechanics,
		\newblock \emph{Elsevier Science}, 2016
		
		\bibitem{Verhagen}
		\newblock Verhagen A.M.W.,
		\newblock  An exactly soluble case of the triangular Ising model in a magnetic field,
		\newblock \emph{Journal of Statistical Physics},  \textbf{15(3) }  (1976), 219-231. DOI: 10.1007/BF01012878
		
		
		\bibitem{Rujan}
    	\newblock Ruj$\acute a$n P. ,        	
    	\newblock Order and disorder lines in systems with competing interactions. III. Exact results from stochastic crystal growth,
		\newblock \emph{Journal of Statistical Physics}, \textbf{34(3-4)} (1984), 615-646. DOI: 10.1007/BF01018562
		
		
		\bibitem{Stephenson}
		\newblock Stephenson J. ,
		\newblock Ising-Model Spin Correlations on the Triangular Lattice. IV. Anisotropic Ferromagnetic and Antiferromagnetic Lattices,
		\newblock \emph{Journal of Mathematical Physics}, \textbf{11(2)} (1970), 420-431. DOI: 10.1063/1.1665155
		
		
		\bibitem{Welberry_Galbraith}
		\newblock Welberry T.R., Galbraith R.,
		\newblock A two-dimensional model of crystal-growth disorder.,
		\newblock \emph{Journal of Applied Crystallography.}, \textbf{6} (1973;), 87-96. DOI: 10.1107/S0021889873008216
		
		\bibitem{Enting}
		\newblock Enting I.G.,
		\newblock Triplet order parameters in triangular and
		honeycomb Ising models,
		\newblock \emph{Journal of Physics A: Mathematical
			and General}, \textbf{10(10)} (1977), 1737-1743. DOI:
		10.1088/0305-4470/10/10/008
		
		\bibitem{Welberry_Miller}
		\newblock Welberry T.R., Miller G.H.,
		\newblock A Phase Transition in a 3D
		Growth-Disorder Model,
		\newblock \emph{Acta Crystallographica Section A:
			Foundations and Advances}, \textbf{A34} (1978), 120-123. DOI: 10.1107/S0567739478000212
		
		\bibitem{Forrester_Baxter}
		\newblock Forrester P.J., Baxter R.J.,
		\newblock Further exact solutions of the
		eight-vertex SOS model and generalizations of the Rogers-
		Ramanujan identities,
		\newblock \emph{Journal of Statistical Physics},  \textbf{38(3-4)} (1985), 435-472. DOI: 10.1007/BF01010471
		
		\bibitem{Jaekel_Maillard_1}
		\newblock Jaekel M.T., Maillard J.M.,
		\newblock A criterion for disorder solutions
		of spin models,
		\newblock \emph{Journal of Physics A: Mathematical and General}, \textbf{18(8)} (1985), 1229-1238. DOI: 10.1088/0305-4470/18/8/023
		
		\bibitem{Baxter_1984}
		\newblock Baxter R.J.,
		\newblock Disorder points of the IRF and checkerboard
		Potts models,
		\newblock \emph{Journal of Physics A: Mathematical and General}, \textbf{17(17)} (1984),L911-L917. DOI: 10.1088/0305-4470/17/17/001
		
		\bibitem{Jaekel_Maillard_2}
		\newblock Jaekel M.T., Maillard J.M.,
		\newblock A disorder solution for a cubic Ising model,
		\newblock \emph{Journal of Physics A: Mathematical and General}, \textbf{18(4)} (1985), 641-651. DOI: 10.1088/0305-4470/18/4/013
		
		\bibitem{Jaekel_Maillard_3}
		\newblock Jaekel M.T., Maillard J.M.,
		\newblock Disorder solutions for Ising and
		Potts models with a field,
		\newblock \emph{Journal of Physics A: Mathematical
			and General}, \textbf{18(12)} (1985),2271-2277. DOI:
		10.1088/0305-4470/18/12/025
		
		\bibitem{Wu1986}
		\newblock Wu F.Y.,
		\newblock Two-Dimensional Ising Model with Crossing and
		Four-Spin Interactions and a Magnetic Field $i ( \pi /2) k T  $,
		\newblock \emph{Journal of Statistical Physics}, \textbf{44(3-4)} (1986), 455-463. DOI: 10.1007/BF01011305
		
		\bibitem{Wu1985}
		\newblock Wu F.Y.,
		\newblock Exact Solution of a Triangular Ising Model in a Nonzero
		Magnetic Field,
		\newblock \emph{Journal of Statistical Physics}, \textbf{40(5-6)} (1985), 613-620. DOI: 10.1007/BF01009892
		
		\bibitem{BDM}
		\newblock Bessis J.D., Drouffe J.M., Moussa P.,
		\newblock Positivity constraints for
		the Ising ferromagnetic model,
		\newblock \emph{Journal of Physics A: Mathematical
			and General}, \textbf{9(12)} (1976), 2105-2124. DOI:
		10.1088/0305-4470/9/12/015
		
		\bibitem{Domb_Green}
		\newblock 
		\newblock Phase Transitions and Critical Phenomena. Vol. 1: Exact Results.
		Domb C., Green H.S. (eds). New York: Academic Press
		\newblock \emph{}, \textbf{} 1972, 506 pp.
		
		\bibitem{Dhar_Maillard}
		\newblock Dhar D., Maillard J.M.,
		\newblock Susceptibility of the checkerboard
		Ising model,
		\newblock \emph{Journal of Physics A: Mathematical and General}, \textbf{18(7)} (1985), L383-L388. DOI: 10.1088/0305-4470/18/7/010
		
		\bibitem{Aguilar_Braun}
		\newblock Aguilar, A., Braun, E.,
		\newblock Exact solution of a general two-dimensional Ising model: The partition function,
		\newblock \emph{Physica A: Statistical Mechanics and Its Applications}, \textbf{170(3)} (1991), 643–662. DOI: 10.1016/0378-4371(91)90011-z
		
		\bibitem{Minlos_Khrapov}
		\newblock Minlos R.A., Khrapov P.V.,
		\newblock Cluster properties and bound
		states of the Yang-Mills model with compact gauge group.I,
		\newblock \emph{Theoretical and Mathematical Physics},  \textbf{61(3)} (1984), 1261-1265. DOI: 10.1007/BF01035013
		
		\bibitem{Khrapov2}
		\newblock Khrapov P.V.,
		\newblock Cluster expansion and spectrum of the transfer
		matrix of the two-dimensional ising model with strong
		external field,
		\newblock \emph{Theoretical and Mathematical Physics}, \textbf{60(1)} (1984), 734-735. DOI: 10.1007/BF01018259
		
		\bibitem{Katrakhov_Kharchenko}
		\newblock Katrakhov V.V., Kharchenko Yu.N.,
		\newblock Two-dimensional fourline
		models of the Ising model type,
		\newblock \emph{Theoretical and Mathematical
			Physics}, \textbf{149(2)} (2006),1545-1558. DOI: 10.1007/s11232-006-0137-y
		
		\bibitem{Khrapov3}
		\newblock Khrapov P.V.,
		\newblock Disorder Solutions for Generalized Ising and
		Potts Models with Multispin Interaction,
		\newblock \emph{Sovremennye informacionnye
			tehnologii i IT-obrazovanie = Modern Information
			Technologies and IT-Education}, \textbf{15(1)} (2019), 33-44. DOI: 10.25559/SITITO.15.201901.33-44
		
		\bibitem{Levenberg}
		\newblock Levenberg K.,
		\newblock A Method for the Solution of Certain Non-Linear
		Problems in Least Squares,
		\newblock \emph{Quarterly of Applied
			Mathematics}, \textbf{2} (1944), 164-168. DOI: 10.1090/qam/10666
		
		
		\bibitem{Perron}
		\newblock Perron O.,
		\newblock Zur Theorie der Matrices,
		\newblock \emph{Mathematische Annalen}, \textbf{64(2)}  (1907), 248-263. DOI: 10.1007/BF01449896
		
		\bibitem{Khrapov4}
		\newblock Khrapov P.V.,
		\newblock Fourier Transform of Transfer Matrices of Plane Ising Models,
		\newblock \emph{Sovremennye informacionnye
			tehnologii i IT-obrazovanie = Modern Information
			Technologies and IT-Education}, \textbf{15(2)} (2019), 306-311. DOI: 10.25559/SITITO.15.201902.306-311
		
		\bibitem{Khrapov5}
		\newblock Khrapov P.V.,
		\newblock Disorder Solutions for Generalized Ising Model with Multispin Interaction,
		\newblock \emph{Sovremennye informacionnye
			tehnologii i IT-obrazovanie = Modern Information
			Technologies and IT-Education}, \textbf{15(2)} (2019), 312-319. DOI: 10.25559/SITITO.15.201902.306-311
		
		
		
		
	\end{thebibliography}
\end{document}